# Astrometric Detection of Earthlike Planets

Michael Shao (Michael.shao@jpl.nasa.gov (Jet Propulsion Laboratory, California Institute of Technology), Geoff Marcy (UC Berkeley), Joseph H. Catanzarite and Stephen J. Edberg (Jet Propulsion Laboratory, California Institute of Technology), Alain Léger (Institut d'Astrophysique Spatiale), Fabien Malbet (Laboratoire d'Astrophysique de l'Observatoire de Grenoble), Didier Queloz (Observatoire de Genève), Matthew W. Muterspaugh (UC Berkeley), Charles Beichman (NExScI), Debra A. Fischer (SFSU), Eric Ford (University of Florida), Robert Olling (University of Maryland), Shrinivas Kulkarni (Caltech), Stephen C. Unwin and Wesley Traub (Jet Propulsion Laboratory, California Institute of Technology)

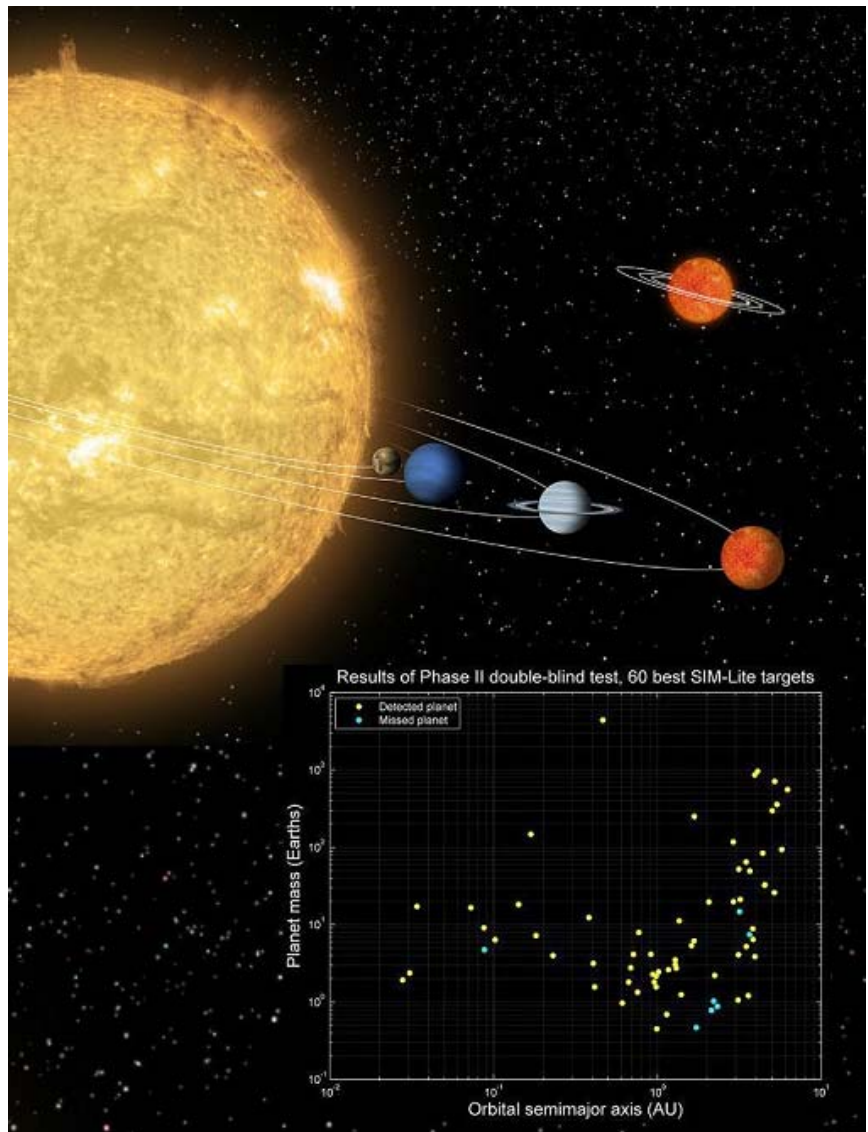

Artist's conception of a binary system with planets and results of a simulation of astrometric planet discoveries from a five year SIM Lite survey of 60 target stars.

## Abstract

Astrometry can detect rocky planets in a broad range of masses and orbital distances and measure their masses and three-dimensional orbital parameters, including eccentricity and inclination, to provide the properties of terrestrial planets. The masses of both the new planets and the known gas giants can be measured unambiguously, allowing a direct calculation of the gravitational interactions, both past and future. Such dynamical interactions inform theories of the formation and evolution of planetary systems, including Earth-like planets.

Astrometry is the only technique technologically ready to detect planets of Earth mass in the habitable zone (HZ) around solar-type stars within 20 pc. These Earth analogs are close enough for follow-up observations to characterize the planets by infrared imaging and spectroscopy with planned future missions such as the James Webb Space Telescope (JWST) and the Terrestrial Planet Finder/Darwin.

Employing a demonstrated astrometric precision of 1 µas and a noise floor under 0.1 µas, SIM Lite can make multiple astrometric measurements of the nearest 60 F-, G-, and K-type stars during a five-year mission. SIM Lite directly tests theories of rocky planet formation and evolution around Sun-like stars and identifies the nearest potentially habitable planets for later spaceborne imaging, e.g., with Terrestrial Planet Finder and Darwin. SIM was endorsed by the two recent Decadal Surveys and it meets the highest-priority goal of the 2008 AAAC Exoplanet Task Force.

## The Trajectory of Exoplanet Research

The field of exoplanet research, since the discovery of planets around nearby stars using radial velocity (RV) measurements by Mayor's and Marcy's groups in 1995, has been growing exponentially, to ~350 at this writing. Planets down to 4 Earth masses have been found in close, four day orbits (Mayor et al. 2009). The spectra of transiting Hot Jupiters have been measured (Knutson et al. 2007, Tinetti et al. 2007, Swain, M. et al. 2008). Young self-luminous planets have been directly imaged around HR 8799 (Marois et al. 2008), Fomalhaut (Kalas et al. 2008), and Beta Pic (Lagrange et al. 2008). With the launch of Kepler, hundreds of transiting Earth-like planets will be found and follow-up observations by HST and JWST will provide us with a huge amount of information on the spectra of jovian- and Neptune- mass planets. Ground based extreme AO (adaptive optics) coronagraphs such as GPI (Gemini Planet Imager) and SPHERE (Spectro-Polarimetric High-contrast Exoplanet Research at the Very Large Telescope) will find many more young self-luminous planets.

The next major advance, and the holy grail of exoplanet research, is the discovery of Earth-like planets in the HZs around nearby stars and their subsequent characterization by direct imaging and spectroscopy. The quest for the discovery of habitable Earth-mass planets has strong scientific as well as public appeal. With astrometry, this quest can start now.

## Science from Astrometric Measurements

**Earth-Analog Characterization**

Astrometric detection of a rocky planet in the HZ (1 AU, scaled) of a host star at sufficient SNR provides its mass and its full, three-dimensional orbit. Moreover, in combination with RV observations, astrometric data will permit the full characterization of all planets, rocky and giant, out to about 5 AU (scaled) for nearby stars. This is an especially intriguing prospect, given that many of the stars that an astrometric mission will examine are already known from RV measurements to have one or more giant planets. Such an inventory of rocky and giant planets around individual nearby stars will offer valuable constraints on the formation and dynamical evolution of planetary systems in general.

**Measuring Masses: A First Step Toward Characterizing Exoplanets**

Determining planet mass by space-based astrometry is clearly necessary in conjunction with direct imaging. Because mass is a fundamental property of planets, the science return of a direct imaging mission would be reduced without astrometry and would suffer significant ambiguities. A dot does not a planet prove, and even a spectrum that reveals some particular molecular constituent (such as methane or carbon dioxide) cannot securely distinguish between a rocky planet and an ice giant. Neither transits (which provide radius) nor spectroscopy (which provides chemical assays of the atmosphere) can solidly discriminate between ice giants and rocky planets. For direct imaging, the uncertain albedos of ice giants and rocky planets will prevent a secure mapping of reflected light fluxes to planet radius, leaving the planetary masses even less secure. Thus, measuring a planet's mass is crucial for interpreting any spectrum of the planet's atmosphere and certainly bears on its habitability. Only by measuring the gravitational effect of the planet on its host star can the planet's mass be measured, and only astrometry can do the job.

An astrometric mission capable of micro-arcsecond (μas) precision is required to empirically characterize Earth-like planets (Exoplanet Task Force 2008). A precise measurement of the planet mass is important when attempting to understand the formation and dynamical evolution of any individual planetary system. Only with accurate measurements of mass, unencumbered by sin (i) ambiguities, can dynamical models be attempted securely. Planet mass may correlate with the mass or chemical composition of the host star and the mass of a particular planet having unexpected properties can spur theoretical research into its formation mechanisms. Theorists will watch for planets having intriguing combinations of mass, orbit, and host star. Table 1 illustrates some of the possibilities for Earthlike planets. Thus, astrometry's ability to characterize individual planet systems, rather than just statistical averages, will help reveal the physical processes by which they formed.

**Table 1**
Planet Properties Enabled By Astrometry plus Visible and Infrared Spectra

| Parameter | Planet Property Derived | Planet Property Implied |
|---|---|---|
| Planet Mass | Density, Surface Gravity, Atmosphere | Likelihood of plate tectonics; atmospheric mass, scale height, lapse rate, and surface pressure and temperature |
| Orbit Semi-Major Axis | Temperature | Potential habitability, liquid water |
| Orbit Eccentricity | Variation of Temperature | Thermal time constant of the atmosphere, mass of the atmosphere |
| Orbit Inclination | Co-Planar Planets? | (System formation mechanism) |
| Orbital Period | Synchronous Rotation? | Potential habitability |

**Astrometric Contributions to the Theory of Rocky Planet Formation**

The theory of the formation of rocky planets and super-Earths, along with their subsequent dynamical evolution, has been developed by Ida and Lin (2008) and Kennedy and Kenyon (2008) and references therein. They start with a collection of thousands of rocky planetesimals having kilometer sizes. Collisions and sticking of these objects allows them to grow and gravitationally perturb each other until final planets emerge. The dynamical effects of the gas and the condensation of ice are included in the calculation. Figure 1 shows the result of a recent simulation by Ida and Lin (2008), displaying the distribution of final planets in a two-parameter space. Remarkably, planets of mass 1 to 30 $M_\oplus$ are predicted to be rare within 1.5 AU. That is, super-Earths and Neptunes are expected to be rare, largely because once inside the ice line they migrate inward quickly, destined to be lost in the star (or parked in a close-in orbit). More-massive rocky cores accrete gas quickly, becoming gas giants. Indeed, our Solar System has no such super-Earths. This prediction may be directly tested by an astrometric mission such as SIM Lite, as planets orbiting within 0.5 to 1 AU, with masses of 5 to 30 $M_\oplus$, will be detected easily.

If the Ida and Lin prediction of a mass desert is contradicted by astrometric observations, the theory must be significantly modified with new physics. Terrestrial planet formation is strongly influenced by gas giant planet formation and orbital evolution. A complete theory of planet formation is needed in order to understand the formation of any one component.

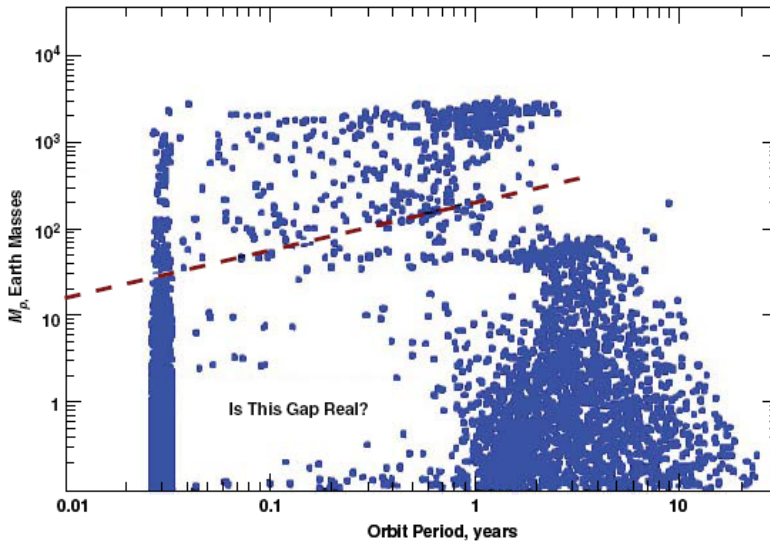

Figure 1. Final planet distribution based from simulations by Ida and Lin (2008).

**Follow-up of Kepler Candidate Earths**

The candidate Earth-like planets revealed by Kepler will raise the question: Are they really Earths? Instead, they could be grazing-incidence giant planets or eclipsing binaries with a brighter third star that dilutes the photometric dimming. Astrometry offers a valuable way to check some of these potential false-positives. Astrometry can detect the motion of the photocenter of the triple-star system, or indeed detect the confused interference fringes from the three stars. With a resolution of 10 mas, astrometric measurements can also interferometrically resolve binaries in the Kepler field. The fractional dimming corresponding to an Earth-sized planet is 1 part in 10,000 around a Sun-sized star. If instead, the dimming is caused by a diluted eclipsing binary, the astrometric shift will be 1 part in 10,000 of the angular separation between the eclipsing binary and the brighter, third star. Ground-based work will detect such "third" stars unless they are within ~0.5 arcsec. For separations of ~0.5 arcsec, the displacement will be 50 µas, easily detectable by an astrometric mission like SIM Lite. For separations of 0.1 arcsec, the displacement will be 10 µas, still detectable by SIM Lite. Thus, SIM Lite offers one of the very few methods to determine the false-positive rate of the Kepler mission.

**Synergy of Astrometry and Direct Imaging of Earths**

In direct imaging of exoplanets, the most important parameter of a coronagraphic observatory is the inner working angle (IWA), the minimum angle for detecting a planet (10 billion times fainter than its parent star) separate from its star's glare. An Earth-Sun system has a maximum angular separation of 0.1 arcsec at 10 pc and 0.05 arcsec at 20 pc. Halving the IWA will increase the number of exo-Earths found by a factor of 8. If we say a planet is detectable if its maximum angular separation is larger than the IWA, then on average 27% of the planets will have maximum separation that is only 10% larger than the IWA and ~50% of the planets will have a maximum separation only 20% larger than the IWA. When a planet is observable only over 10~20% of its orbit, a blind search will need 5~10 images to see the planet once. If only 10% of stars have a terrestrial planet in the HZ, then 50 images, five observations each of 10 stars, on average will yield just one image of one planet. Just the knowledge of the presence of a planet increases the efficiency of direct imaging by a factor of ~10, by avoiding an imaging search of stars known not to have Earths. An astrometric orbit can be used to predict the approximate location to point a direct detection telescope. An exo-Earth detected astrometrically with SNR=6 will have an elliptical error bar 5 years after the mean epoch of the astrometric data. The error in the radial direction of a Earth-Sun clone @10pc will be quite small ~0.03 AU, but the rms error in the azimuthal direction will be ~0.85 radians rms ~0.85 AU. Without any astrometric information the azimuthal error would be +/- pi radians, so an imaging search with an astrometric orbit could be conducted ~3 times faster. In total, an imaging search for exo-Earths would get its first image of an exo-Earth in ~ 1/30 of the time that a "blind" imaging search without astrometric knowledge would take.

If an astrometric orbit is known, confirmation of that orbit by imaging requires just one image. If the planet's orbit is to be determined entirely by imaging, a minimum of three detections are needed to measure the orbit parameters. But the brightness of a planet changes by a factor of a few with orbital phase, and in multi-planet systems it isn't immediately obvious which dot is which planet (especially when the dot only has a SNR=5). A fourth detection of the planet is needed to ensure that all 4 observations were of the same planet. If the planet is outside the IWA only over 20% of its orbit, ~20 images will be needed to see the planet 4 times and many orbital parameters will still be almost impossible to measure, such as orbit inclination and eccentricity.

In summary, if the fraction of stars with an Earth in the HZ is ~10%, an astrometric precursor would speed up the 1$^{st}$ image of an exo-Earth by roughly a factor of 30 and a subsequent "accurate" orbit for follow up spectroscopic observations would be a factor ~10 faster because only 1 image combined with 5 yrs of astrometric data is needed to narrow the azimuthal error bar.

**Planets Orbiting Binary Stars**

Astrometry is also capable of detecting planets orbiting binary stars. About half of the stars in the solar neighborhood are in binary systems so this represents a significant fraction of the potentially habitable worlds. If an astrometric mission finds a significant number of Earthlike planets around binary stars, this might strongly influence the architecture of a direct imaging mission to one that is able to directly image planets around double stars. About 1/2 of the coronagraphic architectures currently being studied are able to image planets around double stars. The external occulter is an example of an architecture that isn't compatible with double stars.

**Lifting the Mass-Inclination Ambiguity in RV Planets**

The large majority of known exoplanets have ambiguous masses: their unknown inclination angles permit only lower-limit mass estimations. This mass ambiguity disappears in the astrometric determinations of dynamical mass, since astrometry determines the system inclination angles. Specific targets of interest where unambiguous masses are important include those with multiple planets where planet–planet interactions might become significant. The inclination ambiguity does not strongly affect the derived distributions of planet masses. However, for individual planetary systems, it is important to know the full specification of masses and orbits to compare with the structure of the systems predicted by theorists. Ida and Lin (2004) demonstrate how the knowledge of the unambiguous dynamical masses guides the development of the theory. Unknown inclination angles (and hence masses) degrade the quality of the comparison between observation and theory, especially when planet-planet interactions are significant.

### Astrometric Detection of Exo-Earths

The semi-amplitude α of the angular wobble of a star of mass $M_*$ and distance $D$ due to a planet of a given mass $M_p$ orbiting with a semi-major axis $a$ is given by:

$$\alpha = 3.00 \frac{M_\odot}{M_*} \frac{M_p}{M_\oplus} \frac{a}{(1 \text{ AU})} \frac{(1 \text{ pc})}{D} \text{ μas}$$

The benchmark case is an Earth-mass planet orbiting 1 AU from a solar-mass star located 10 pc away. For such a planet, the astrometric semi-amplitude α is 0.3 μas. Standard time-series analysis provides a quantitative measure of the increasing signal-to-noise ratio that accrues with an increasing number of observations. Detection of a periodic signal with 1% false alarm probability requires an end of mission SNR of ~5.8. Here, we define

$$\text{SNR} = \alpha N^{1/2} / \sigma,$$

where α is the semi-major axis of the astrometric signal, σ is the single epoch noise and N is the number of epochs at the end of the mission (Catanzarite et al. 2006).

**Earths Are Detectable Despite the Presence of Other Planets**

A significant question is whether an astrometric mission can detect Earth-mass planets in the presence of the astrometric noise contributed by other planets, both other terrestrials and gas giants. Planets and asteroids located both inward and outward of the HZ create a forest in the power spectrum, making the detection of 1 Earth-mass planets difficult. Giant planets orbiting well outside 1 AU cause an astrometric curvature partially absorbed (incorrectly) in the solution for proper motion.

In a comprehensive 'double-blind' test of astrometric planet finding and characterization with SIM Lite, 60 planet-search target stars were perturbed with simulated planetary systems that were generated by five teams of dynamicists. Astrometric and RV observation data were generated for each mock system with realistic errors and cadence. The astrometric and velocity data were analyzed by four teams without knowledge of the input planets. Planets with astrometric amplitude greater than $5.8\sigma/N^{1/2}$ are detectable, where $\sigma$ is the single-measurement error and N is the number of observations (Catanzarite et al. 2006). Of the 70 planets that were detectable, 63 were found, for an overall completeness of 90%. Of the 18 terrestrial planets in the HZ that were detectable, 17 were found, for a completeness of 94%. One false detection was registered, and it was not in the HZ; the overall reliability was 99%. Masses of most detected planets were estimated to better than 8%, and periods to better than 1%. Results of the double-blind test are shown in Table 2 and on the front cover.

**Table 2**
Results of the Double-Blind Exo-Earth Detection Test

|  | Detectable | Detected | Completeness |
|---|---|---|---|
| Planets in the habitable zone | 22 | 21 | 95% |
| Terrestrial planet | 43 | 37 | 86% |
| Terrestrial planet in the habitable zone | 18 | 17 | 94% |
| Overall | 70 | 63 | 90% |

A major conclusion of the double-blind study is that the presence of additional planets in a planetary system causes almost no degradation in the detectability of Earth-mass planets. This double-blind test demonstrates that SIM Lite is capable of detecting Earth-like planets embedded in multiple-planet systems around the 60 most suitable nearby planet search target stars, and of accurately determining their masses and orbits.

**Earth Search Yields**

The exact expenditure of SIM Lite time spent searching for other Earths remains open to discussion. The number of target stars studied is strongly correlated with the percentage of mission time devoted to the search and the exoplanet mass limit desired. Raising the mass limit searched rapidly increases the number of stars that can be searched per unit mission time.

To survey the nearest 64 stars to a threshold of 1 $M_\oplus$ at 1 AU (scaled to stellar luminosity) takes 40 percent of the nominal SIM Lite mission time, with a lifetime of five years. The exact expenditure of SIM Lite time spent searching for other Earths remains open to discussion. An alternative is to search 135 stars to a sensitivity of 2 Earth mass. A more detailed observation plan will be developed after initial results from the Kepler mission are available.

## Instrumental and Astrophysical Noise

To detect an Earth in an Earth-Sun system at 10 pc means a space astrometric instrument must have an end-of-mission narrow angle precision of ~0.05 μas. For a 6 m baseline stellar interferometer, this implies the ability to control systematic errors to measure the optical path of starlight with ultimate precision of ~1.5 picometers (Shao and Nemati 2009; see also http://planetquest.jpl.nasa.gov/SIM/files/Chapter-19-LR.pdf).

In addition to instrumental noise, one must also consider astrophysical noise sources. A slightly large sunspot (of 0.1% of the area of the solar disk) will bias the photo-center of the Sun at 10 pc by ~0.25 μas. That same sunspot will produce ~1 m/s RV bias. The astrometric amplitude of an Earth at 1 AU and 10 pc is 0.3 μas and the RV amplitude is 0.1 m/s. Roughly speaking, astrophysical noise from stellar activity is ~10 times larger for RV than for astrometry for planets with ~1 yr period. The astrometry and RV bias of star spots have a time scale of 1~2 weeks, meaning that this noise can average down no faster than $N_{weeks}^{1/2}$. RV and astrometry have opposite sensitivities relative to orbital period. Short period planets are more easily detected by RV and long period planets more easily by astrometry.

## Summary

The next major advance in exoplanet research is the detection of Earth mass planets in the HZ of nearby stars. The technology for a space astrometric observatory capable of detecting Earths in the HZ of nearby stars has been demonstrated in the laboratory. A survey of 60~100 nearby stars is needed to ensure that a reasonable number of Earths would actually be detected, and this could be done with a five year space astrometry mission. Only astrometry can measure the mass of the planet unambiguously. The mass of the planet determines whether the planet will hold on to an atmosphere or whether the atmosphere is composed of molecules like $CO_2$, $H_2O$, $O_2$, etc. or is primarily $H_2$, like Neptune or Jupiter. An astrometry mission is synergistic with direct imaging, enabling a smaller telescope or interferometer for direct detection, and if many Earths are found around binary stars, favoring some types of direct imaging architectures over others. Splitting the discovery and characterization of other Earths between two sequential missions, namely an easier indirect discovery program like SIM Lite and a later imaging and spectroscopy program, e.g., Terrestrial Planet Finder/Darwin, allows for technology to be developed for the latter with specified detection thresholds. Such a two-step approach offers methodical, cumulative, and complementary information about other Earth-like planets, Neptunes, and jovian planets, and their systems.

## Acknowledgments

The research described in this paper was carried out at the Jet Propulsion Laboratory, California Institute of Technology, under contract with the National Aeronautics and Space Administration. Copyright 2009 California Institute of Technology. Government sponsorship acknowledged.